\documentclass[12pt]{article}
\makeatletter  

\@addtoreset{equation}{section} \makeatother

\newcommand{\hs}{\hspace*{0.5cm}}
\newcommand{\vs}{\vspace*{0.5cm}}
\newcommand{\be}{\begin{equation}}
\newcommand{\ee}{\end{equation}}
\newcommand{\bea}{\begin{eqnarray}}
\newcommand{\eea}{\end{eqnarray}}
\newcommand{\ben}{\begin{enumerate}}
\newcommand{\een}{\end{enumerate}}
\newcommand{\nn}{\nonumber}
\newcommand{\crn}{\nonumber \\}

\newcommand{\al}{\alpha}

\newcommand{\ga}{\gamma}
\newcommand{\va}{\varphi}
\newcommand{\om}{\omega}
\newcommand{\pa}{\partial}
\newcommand{\fr}{\frac}
\newcommand{\bc}{\begin{center}}
\newcommand{\ec}{\end{center}}

\newcommand{\de}{\delta}

\def\lappeq{\mathrel{\rlap{\raise.5ex\hbox{$<$}}
{\lower.5ex\hbox{$\sim$}}}}
\begin{document}

\bc {\Large Atomic parity violation in the economical 3-3-1 model}\\
\vspace*{1cm}

{\bf P. V. Dong, H. N. Long and D. T. Nhung}\\

\vspace*{0.5cm}

{\it Institute of Physics, VAST, P. O. Box 429, Bo Ho, Hanoi
10000, Vietnam}\\

\ec

\begin{abstract}
The deviation $\de Q_{\mathrm{W}}$ of the weak charge from its
standard model prediction due to the mixing of the $W$ boson with
the charged bilepton $Y$ as well as of the $Z$ boson with the
neutral $Z'$ and the real part of the non-Hermitian neutral
bilepton $X$ in the economical 3-3-1 model is established.
Additional contributions to the usual $\de Q_\mathrm{W}$
expression in the extra $\mathrm{U}(1)$ models and the left-right
models are obtained. Our calculations are quite different from
previous analyzes in this kind of the 3-3-1 models and give the
limit on mass of the $Z'$ boson, the $Z-Z'$ and $W-Y$ mixing
angles with the more appropriate values: $M_{Z'}
> 564\ \mathrm{GeV},\ -0.018<\sin \va < 0$ and $|\sin \theta| < 0.043$.

\vs


\end{abstract}

\section{Introduction}
\label{Intro}

Recent neutrino  experimental results establish the fact that
neutrinos have masses and the standard model (SM) must be
extended. Among the beyond-SM extensions, the models based on the
$\mathrm{SU}(3)_C\otimes \mathrm{SU}(3)_L \otimes \mathrm{U}(1)_X$
(3-3-1) gauge group \cite{ppf,flt} have some intriguing features:
First, they can give partial explanation of  the generation number
problem. Second,  the third quark generation has to be different
from the first two, so this leads to the possible explanation of
why top quark is uncharacteristically heavy.

In one of the 3-3-1 models, the scalar sector is minimal with just
two Higgs triplets; hence it has been called the economical 3-3-1
model~\cite{ponce,haihiggs}. The general  Higgs sector of this
model is very simple (economical) and consists of three physical
scalars (two neutral and one charged) and eight Goldstone bosons -
the needed number for massive ones~\cite{higgseconom}.

It is well known that the atomic parity violation (APV) in Cesium
is one of the most useful sources of information on the still
unknown parameters of the SM and on possible physics beyond the
SM. In Ref.~\cite{longtrung}, the APV in Cesium has been used to
fix parameters in the usual 3-3-1 models. In these models and in
similar considered in~\cite{alta,lalu}, the parity violation is
mediated through the neutral $Z$ and $Z'$ gauge bosons. However,
in the economical 3-3-1 model, the APV effect gets additional
contribution from the real part of neutral non-Hermitian gauge
boson ($X^0+X^{0*}=\sqrt{2} W_4$). In the framework of the
economical 3-3-1 model, the APV has been considered \cite{ponc1},
however neglected the contribution from the real part $W_4$ as
well as from the $W - Y$ mixing.

The aim of the present work is to calculate a contribution from
the above mentioned  mixings, namely, the mixing among $W,\ Y$ and
among  $ W_4,\ Z,\ Z'$.

The rest of this  paper is organized as follows: In Section
\ref{model}, we briefly introduce necessary  elements of the
model, the mixtures in the gauge boson sector.
 Sec. \ref{sec1} is devoted to calculating the new gauge boson contributions
to the weak charge. We summarize our result and make conclusions
in the last section - Sec.\ref{concl}.

\section{\label{model} Mixing among SM and new gauge
bosons in the economical 3-3-1 model}
 The particle content in this model, which is anomaly
free, is given as follows: \be \psi_{aL} = \left(
               \nu_{aL}, l_{aL}, N_{aL}
\right)^T \sim (3, -1/3),\hs l_{aR}\sim (1, -1),
  \label{l2}
\ee where $a = 1, 2, 3$ is a  family index and the right-handed
neutrino denoted by $ N_L \equiv (\nu_R)^c$. \bea
 Q_{1L}&=&\left( u_{1L},  d_{1L}, U_L \right)^T\sim
 \left(3,1/3\right),\hs Q_{\al L}=\left(
  d_{\al L},  -u_{\al L},  D_{\al L}
\right)^T\sim (3^*,0),\hs \al=2,3,\crn u_{a
R}&\sim&\left(1,2/3\right),\hs d_{a R} \sim
\left(1,-1/3\right),\hs U_{R}\sim \left(1,2/3\right),\hs D_{\al R}
\sim \left(1,-1/3\right).\eea
 Here, the values in the parentheses denote quantum
numbers based on the $\left(\mbox{SU}(3)_L,\mbox{U}(1)_X\right)$
symmetry. Electric charges of the exotic $U$ and $D_\al$ quarks
are the same as of the usual quarks, i.e., $q_{U}=2/3$,
$q_{D_\al}=-1/3$. The electric charge operator is given in the
form\be Q=T_3-\fr{1}{\sqrt{3}}T_8+X,\label{eco}\ee where $T_i$
$(i=1,2,...,8)$ and $X$ stand for $\mbox{SU}(3)_L$ and
$\mbox{U}(1)_X$ charges, respectively. Thus, the SM electric
charge operator, $Q=T_3+Y_{\mathrm{w}} $, is embedded into the
larger group via weak hypercharge extension:\be
Y_{\mathrm{w}}=-\fr{1}{\sqrt{3}}T_8+X.\label{hch}\ee

The spontaneous symmetry breaking in the model under consideration
is achieved via the two Higgs scalar triplets
\cite{ponce,haihiggs} \be \chi=\left(  \chi^0_1, \chi^-_2,
\chi^0_3 \right)^T \sim \left(3,-1/3\right), \hs \phi=\left(
\phi^+_1,  \phi^0_2, \phi^+_3 \right)^T\sim \left(3,2/3\right)\ee
with vacuum expectation values (VEVs) given by \be
\langle\chi\rangle=\fr{1}{\sqrt{2}}\left(  u,  0,  \om \right)^T,
\hs \langle\phi\rangle =\fr{1}{\sqrt{2}}\left(
  0,  v,  0 \right)^T.\ee

Because the field $\chi^0_1$ has lepton number equal to two units,
the VEV $u$ is a kind of the lepton number violating parameters.
To keep a consistency  with the low energy phenomena, the VEVs
must be satisfied by the
constraints~\cite{haihiggs,higgseconom}:\be u^2 \ll  v^2 \ll
\om^2. \label{vevcons} \ee

The VEV $\om$ gives mass for the exotic quarks $U$ and $D_\al$,
$u$ gives mass for $u_1, d_{\al}$ quarks, while $v$ gives mass for
$u_\al, d_{1}$ and all ordinary leptons. At the tree level, the
mass matrix for the up-quarks has one massless state and in the
down-quark sector, there are two massless ones. With the minimal
scalar content, radiative correction, however,  will modify the
quark spectrum. The radiative pattern for the quark masses in the
economical 3-3-1 model is out of the scope of this work and will
be presented elsewhere.

The covariant derivative of a triplet is \be D_\mu =
\pa_\mu-igT_iW_{i\mu}-ig_X T_9 X B_\mu \equiv
\pa_\mu-i\mathcal{P}_\mu,\ee where
$T_9=\mathrm{diag}(1,1,1)/\sqrt{6}$ is fixed by $\mbox{Tr}(T_{i}
T_{j})=\delta_{ij}/2$ $(i,j=1,2,...,9)$. The $\mathcal{P}_\mu$
matrix can be rewritten in a convenient form \bea
{\mathcal{P}}_\mu=
\fr{g}{2}\left(%
\begin{array}{ccc}
  W_{3\mu}+\fr{1}{\sqrt{3}}W_{8\mu}+t\sqrt{\fr 2
3}XB_\mu & \sqrt{2} W^+_{0\mu} & \sqrt{2}X^0_{0\mu} \\
  \sqrt{2}W^-_{0\mu} & -W_{3\mu}+\fr{1}{\sqrt{3}}W_{8\mu}+
t\sqrt{\fr 2 3}X B_\mu & \sqrt{2}Y^-_{0\mu} \\
  \sqrt{2}X^{0*}_{0\mu} & \sqrt{2}Y^+_{0\mu} & -\fr{2}{\sqrt{3}}W_{8\mu}
+t\sqrt{\fr 2 3}X B_\mu\\
\end{array}%
\right),\nn \eea where $t\equiv
g_X/g=3\sqrt{2}s_W/\sqrt{4c^2_W-1}$ with $s_W \equiv e/g$. Here,
the combinations \be W^{\pm} _{0} \equiv \fr{W_{1}\mp
iW_{2}}{\sqrt{2}},\hs Y^\mp_{0} \equiv \fr{W_{6}\mp
iW_{7}}{\sqrt{2}}, \hs X^0_{0} \equiv \fr{W_{4}-iW_{5}}{\sqrt{2}}
\label{chags} \ee have the defined charges (indicated by the
superscripts $^{+,-,0}$) under the generators of the
$\mathrm{SU}(3)_L$ group. The subscript $_0$ which is associated
with the fields $W_0$, $X_0$ and $Y_0$ implies that they are not
mixing and {\it unphysical} as shown below. For the sake of
convenience in further reading, we remind the reader that $W_{4}$
and $W_{5}$ are pure real and imaginary parts of $ X_0^0$ and $
X^{0*}_0$, respectively \be W_{4} = \fr{1}{\sqrt{2}} ( X^0_{0} +
X^{0*}_{0}),\hs W_{5} = \fr{i}{\sqrt{2}} ( X^0_{0} -
X^{0*}_{0}).\label{w4xx}\ee

In the model, the SM charged boson $W_0$ is mixing with the new
bilepton $Y_0$ via the following mass Lagrangian \cite{haihiggs}
\be
{\mathcal{L}}^{\mathrm{C}}_{\mathrm{mass}}=\fr{g^2}{4}(W_0,Y_0)\left(%
\begin{array}{cc}
  u^2+v^2 & u\om \\
  u\om & \om^2+v^2 \\
\end{array}%
\right)\left(%
\begin{array}{c}
  W_0 \\
  Y_0 \\
\end{array}%
\right).\nn\ee Therefore, the physical fields in this sector are
obtained in terms of a mixing angle $t_\theta=u/\om$ by \bea W&=&
c_\theta W_{0}- s_\theta Y_{0},\crn Y&=& s_\theta W_{0}+c_\theta
Y_{0},\label{cct}\eea where $s_\theta \equiv \sin\theta$,
$t_\theta \equiv \tan\theta$, and so forth. These bosons, $W$ and
$Y$, have the masses defined, respectively,  by \be
M^2_{W}=\fr{g^2v^2}{4},\hs
M^2_{Y}=\fr{g^2}{4}(u^2+v^2+\om^2).\label{massy}\ee Thus, under
the constraints (\ref{vevcons}), the $W$  is the SM-like gauge
boson and $v\approx v_{\mathrm{weak}}=246\ \mathrm{GeV}$.

In the neutral sector, under conservation of the electric charge,
the physical photon field is independent of the VEVs and
associated with the electric charge operator (\ref{eco}) in a
general form \cite{haihiggs,dolong} \be A = s_W
W_{3}+c_W\left(-\fr{t_W}{\sqrt{3}}
W_{8}+\sqrt{1-\fr{t^2_W}{3}}B\right).\label{photon}\ee The photon
field is not mixing with other neutral ones. By embedding the
electric charge operator (\ref{eco}) into the SM symmetry, the SM
neutral boson $Z_0$ is defined  orthogonally to $A$ as follows\be
Z_{0} = c_W W_{3}-s_W\left(-\fr{t_W}{\sqrt{3}}
W_{8}+\sqrt{1-\fr{t^2_W}{3}}B\right).\ee The new neutral boson
$Z'_0$ arises from the extra symmetry for the weak hypercharge
(\ref{hch}) in terms of \be Z'_{0} = \sqrt{1-\fr{t^2_W}{3}}
W_{8}+\fr{t_W}{\sqrt{3}}B.\ee Here, the $Z'_0$ is defined
orthogonally to the photon field in the extra part associated with
$Y_{\mathrm{w}}$.

Similarly to the photon field, the imaginary part $W_5$ of the
non-Hermitian gauge boson $X_0$ is decoupled; whereas, its real
part mixes with $Z_0$ and $Z'_0$. This can be expressed via the
mass Lagrangian as follows \cite{haihiggs} \bea
{\mathcal{L}}^{\mathrm{N}}_{\mathrm{mass}}=\fr{1}{2}M^2_{X} W^2_5+
\fr{1}{2}V^T_0 M^2 V_0,\eea where \bea M^2_X &\equiv&
\fr{g^2}{4}\left(u^2+\om^2\right),\label{massx} \\ V_0 &\equiv&
\left(Z_0,Z'_0, W_4\right)^T, \label{v00}
\\ M^2 &\equiv& \left(%
\begin{array}{c c c}
  \fr{g^2(u^2 + v^2)}{4c^2_W} &\ \fr{g^2(c_{2W}u^2-v^2)}{4c^2_W
  \sqrt{3-4s^2_W}} \ & \fr{g^2 u \om}{2c_W} \\ \\
  \fr{g^2(c_{2W}u^2-v^2)}{4c^2_W\sqrt{3-4s^2_W}}\hs &\hs
  \fr{g^2(v^2+4c^4_W\om^2+c^2_{2W}u^2)}{4c^2_W(3-4s^2_W)}\hs & \hs
  -\fr{g^2 u \om}{2c_W\sqrt{3-4s^2_W}} \\ \\
  \fr{g^2 u \om}{2c_W} & -\fr{g^2u \om}{2c_W\sqrt{3-4s^2_W}} & M^2_X \\
\end{array}%
\right).\label{nmass}\eea It is to be noted that the  mass matrix
(\ref{nmass}) {\it contains} an exact eigenvalue equal to $M^2_X$
with the eigenstate given by~\cite{haihiggs} \bea W'_{4} &=&
s_{\theta'} Z_{0}+
c_{\theta'}\left[t_{\theta'}\sqrt{4c^2_W-1}Z'_{0}+
\sqrt{1-t^2_{\theta'}(4c^2_W-1)}W_{4}\right],\label{htmix}\\
s_{\theta'}&\equiv &
\fr{t_{2\theta}}{c_W\sqrt{1+4t^2_{2\theta}}}.\nn \eea The $W'_4$
and $W_5$ have the same mass,  therefore their combination has to
be identified with {\it physical} neutral gauge boson: \be X^0
\equiv \fr{W'_{4} - i W_{5}}{\sqrt{2}} \label{chbln} \ee with mass
$M_X$. This is a crucial difference between our treatment and that
in Ref.~\cite{ponce}. It is worth emphasizing that, in
Ref.~\cite{ponce}, the real part $W_4$  and the imaginary part
$W_5$ are treated as the physical fields. It is easy to see that
$X^0$ is the bilepton gauge boson.

Comparing (\ref{htmix}) with (\ref{photon}), we see that $W'_{4}$,
by structure, is quite similar to $A$; and the mixing angle, in
this case,  $\theta'$ has the origin from the lepton-number
violating ($\propto u$). It was shown that~\cite{haihiggs} the
$W'_{4}$ decouples from two gauge vectors: \bea
\mathcal{Z}&=&c_{\theta'} Z_{0}-
s_{\theta'}\left[t_{\theta'}\sqrt{4c^2_W-1}Z'_{0}+
\sqrt{1-t^2_{\theta'}(4c^2_W-1)}W_{4}\right],\crn
\mathcal{Z}'&=&\sqrt{1-t^2_{\theta'}(4c^2_W-1)}Z'_{0}-
t_{\theta'}\sqrt{4c^2_W-1}W_{4}.\eea Now there remains only a
mixing between $\mathcal{Z}$ and $\mathcal{Z}'$ through an angle
$\va$ \bea t_{2\va}&=&\sqrt{(3-4s^2_W)(1+4t^2_{2\theta})}
\left\{[c_{2W}+(3-4s^2_W)t^2_{2\theta}]u^2-
v^2-(3-4s^2_W)t^2_{2\theta}\om^2\right\}\crn &&\times
\left\{[2s^4_W-1+(8s^4_W-2s^2_W-3) t^2_{2\theta}]u^2-
[c_{2W}+2(3-4s^2_W)t^2_{2\theta}]v^2\right.\crn&&+\left.[2c^4_W+
(8s^4_W+9c_{2W})t^2_{2\theta}]\om^2\right\}^{-1}.\eea
 In terms of
$\va$, the remaining physical fields are given by
 \bea Z &=& c_\va \mathcal{Z}-s_\va
\mathcal{Z}',\crn Z' &=& s_\va \mathcal{Z}+c_\va \mathcal{Z}',\eea
which are signified via the masses \bea
M^2_{Z}&=&[2g^{-2}(3-4s^2_W)]^{-1}\left\{c^2_W(u^2+\om^2)+v^2\right.
\crn &&-\left.\sqrt{[c^2_W
(u^2+\om^2)+v^2]^2+(3-4s^2_W)(3u^2\om^2-
u^2v^2-v^2\om^2)}\right\},\crn
M^2_{Z'}&=&[2g^{-2}(3-4s^2_W)]^{-1}\left.\{c^2_W(u^2+\om^2)+v^2\right.
\crn&& \left. +\sqrt{[c^2_W
(u^2+\om^2)+v^2]^2+(3-4s^2_W)(3u^2\om^2
-u^2v^2-v^2\om^2)}\right\},\nn\eea to the SM-like $Z$ gauge boson
and the exotic of the new physics $Z'$, respectively. Finally,
such mixings can be demonstrated via a mixing matrix $\mathrm{U}$
relating $V_0$ to the physical fields $V = (Z,Z',W'_4)^T$ as
follows: \be V=\mathrm{U}V_0,\hs M^2_V\equiv
\mathrm{diag}(M^2_Z,M^2_{Z'},M^2_X)=\mathrm{U} M^2
\mathrm{U}^T,\ee
\be \mathrm{U}\equiv \left(%
\begin{array}{ccc}
  c_\va & -s_\va & 0 \\
  s_\va & c_\va & 0 \\
  0 & 0 & 1 \\
\end{array}%
\right)\left(%
\begin{array}{ccc}
  c_{\theta'} & -s_{\theta'}t_{\theta'}\sqrt{4c^2_W-1} & -s_{\theta'}\sqrt{
  1-t^2_{\theta'}(4c^2_W-1)}\\
0 & \sqrt{1-t^2_{\theta'}(4c^2_W-1)}& -
t_{\theta'}\sqrt{4c^2_W-1}\\
  s_{\theta'} & c_{\theta'}t_{\theta'}\sqrt{4c^2_W-1} &
c_{\theta'}\sqrt{1-t^2_{\theta'}(4c^2_W-1)} \\
\end{array}%
\right).\ee

It is important to note that, due to the $W_0-Y_0$ and
$Z_0-Z'_0-W_4$ mixings, the $\rho=1+\Delta \rho$ parameter,
defined by the relation \be c^2_W=\fr{M^2_W}{\rho M^2_Z}
\label{drho}\ee receives a contribution $\Delta \rho_{\mathrm{m}}$
given by: \bea \Delta
\rho_{\mathrm{m}}&=&\left[\mathrm{U}^2_{11}+\fr{\mathrm{U}^2_{21}M^2_{Z'}+
\mathrm{U}^2_{31}M^2_X}{M^2_Z}\right]
\left[c^2_\theta+s^2_\theta\fr{M^2_Y}{M^2_W}\right]^{-1}-1\\
&\simeq& (c_{\theta'}s_\va)^2
\left[\fr{M^2_{Z'}}{M^2_Z}-1\right]+s^2_{\theta'}\left[\fr{M^2_X}{M^2_Z}-1\right]
-s^2_\theta\left[\fr{M^2_Y}{M^2_W}-1\right].\label{rho} \eea

Due to the $W_0-Y_0$ mixing, the charged currents in this model
contain the unnormal part given in ~\cite{haihiggs}. Therefore,
the Fermi coupling constant which is derived from effective
interactions takes  the form (at the tree level) \be
\fr{G_{\mathrm{F}}}{\sqrt{2}}=\fr{g^2}{8}\left(\fr{c^2_\theta}{M^2_W}
+\fr{s^2_\theta}{M^2_Y}\right).\ee With the help of (\ref{drho}),
we have a good approximation as follows \be s^2_W
c^2_W\simeq\left(\fr{\pi\al}{\sqrt{2}G_{\mathrm{F}} M^2_Z}\right)
\fr{c^2_\theta}{\rho}.\ee Hence, we see that the modifications to
the $\rho$ parameter and the gauge coupling constant $g\rightarrow
g c_\theta$ affect such a relation of the SM between the $s_W$ and
the basic electroweak parameters such as $\al$, $G_F$ and $M_Z$.
This produces a shift on $s^2_W$ already at the tree level given
by \be \de(s^2_W)\simeq -\fr{s^2_W c^2_W}{c_{2W}}(\Delta
\rho_{\mathrm{m}}+s^2_\theta).\label{modii}\ee This modification
must be appropriately taken into account in evaluating the
deviations of experimental quantities given in this model from
their SM predictions. The quantity $\Delta \rho_{\mathrm{m}}$ adds
to the other possible terms contributing to the $\rho$ parameter,
such as $\Delta \rho_T= \al T$
coming from the leading electroweak radiative correction (for
details, see~\cite{il}), therefore, in this case $\Delta
\rho_{\mathrm{m}} \rightarrow \Delta \rho = \Delta
\rho_{\mathrm{m}} + \Delta \rho_T$. As a typical example of atomic
parity violation, next, we will consider the deviation of the weak
charge in this model.

\section{\label{sec1} Deviation of the weak charge $\de Q_{\mathrm{W}}$}
Basic to the analysis of the experiments on parity violation in
atoms is electron-quark effective Lagrangian: \be
\mathcal{L}_{\mathrm{eff}}=\fr{G_{\mathrm{F}}}{\sqrt{2}}\left(\overline{e}\ga_\mu
\ga_5 e\right)\left(C_{1u} \overline{u} \ga^\mu u +C_{1d}
\overline{d} \ga^\mu d\right).\label{leef}\ee The parity violating
effect for the vector-coupled electrons and axially-coupled quarks
is strongly suppressed because of the smaller vector coupling of
electrons in the SM and of its dependence on spins rather than
charges. We shall not discuss this effect here, which is very
small in a heavy atom. The experimental results are usually given
in terms of the related so-called weak charge: \be Q_{\mathrm{W}}=
-2\left[C_{1u} (2Z+N)+C_{1d} (Z+2N)\right],\label{wcharge}\ee
where $Z$ and $N$ are the number of protons and of neutrons in the
considered atom, respectively.

Contribution to the SM  weak charge, in the models with an extra
U(1) and in the left-right symmetric models, by only one new
neutral gauge boson $Z'$ has been considered in
Ref.\cite{alta,lalu}. In the model under consideration, {\it all}
new gauge bosons such as $Z'$, $X$ and $Y$ will correct
$Q_{\mathrm{W}}$ from its SM prediction by the deviation $\de
Q_{\mathrm{W}} = Q_{\mathrm{W}}-Q^{\mathrm{SM}}_{\mathrm{W}} $.
This deviation will be expressed in terms of the SM three
parameters mentioned before: $\al$, $G_{\mathrm{F}}$, and $Z$
boson mass $M_Z$ which have been obtained from the experiments
with fair accuracy \cite{pdg}. We will show that it consists of
the sum of four terms: \ben \item The first term is proportional
to the shift $\Delta \rho$ of the $\rho$ parameter which is
originated from the $W_0-Y_0$ and $Z_0-Z'_0-X_0$ mixings and from
the leading radiative corrections.\item The second and third terms
are proportional to the mixing angles $\va$ and squared
$\theta$.\item The last term is proportional to the squared mass
ratio $M^2_Z/M^2_{Z'}$.\een As shown \cite{haihiggs} and in the
following, all these contributions are small, thus their higher
orders may be neglected.

Now we turn to the deviation of the weak charge $\de
Q_{\mathrm{W}}$. First, in the base of $V_0$, the neutral currents
are obtained via the interaction Lagrangian \be \mathcal{L}=V^T_0
H,\hs H \equiv (J,J',J_4)^T,\label{ggg}\ee with \bea J_{\mu}
&=&\fr{g}{c_W}\left(v_f \overline{f}\ga_\mu f+a_f \overline{f}
\ga_\mu \ga_5 f\right),\crn J'_{\mu} &=&\fr{g}{c_W}\left(v'_f
\overline{f}\ga_\mu f+a'_f \overline{f} \ga_\mu \ga_5
f\right),\crn J_{4\mu} &=& 0.\nn\eea Here $f$ represents the
generic fermion, and the vector and axial-vector couplings $v_f$,
$v'_f$, $a_f$ and $a'_f$ relevant to this analysis are given in
Table \ref{tabe}. The second and the third columns correspond to
the case when $u$ and $d$ quarks belong to triplet and
antitriplet, respectively.
\begin{table}[htb]
\caption{\label{tabe} Vector and axial-vector couplings relevant
for APV in the SM and in the economical model (EcM) where $u$ and
$d$ quarks belong to triplet/antitriplet .}
\begin{tabular}{@{}l|l|l}
SM & EcM  $u, d \in  $ triplet    & EcM, $u, d \in $ antitriplet
\\ \hline $a_e=\fr{1}{4}$ &
$a'_e=-\fr{1}{4}\left(3-4s^2_W\right)^{-\fr 1
2}$ & $a'_e=-\fr{1}{4}\left(3-4s^2_W\right)^{-\fr 1 2}$ \\
$v_u=\fr{1}{4}-\fr{2}{3}s^2_W$ &
$v'_u=\left(\fr{1}{4}+\fr{1}{6}s^2_W\right)\left(3-4s^2_W\right)^{-\fr
1 2}$ & $v'_u=-\left(\fr 1 4 - \fr 2 3
s^2_W\right)\left(3-4s^2_W\right)^{-\fr 1 2}$ \\
 $v_d=-\fr 1 4
+\fr 1 3 s^2_W$ & $v'_d= \fr{1}{12}\left(3-4s^2_W\right)^{\fr 1
2}$ & $v'_d = -\left(\fr 1 4 -\fr 1 6
s^2_W\right)\left(3-4s^2_W\right)^{-\fr 1 2 }$
\end{tabular}\\[6pt]
\end{table}

Finally, in the physical basis $V$, the Lagrangian (\ref{ggg})
becomes \be \mathcal{L}=V^T U H.\label{ggg1}\ee At the low-energy
limit, the propagator for the field $V$ can be approximated by \be
\fr{-ig^{\mu\nu}}{q^2-M^2_V}\rightarrow
\fr{ig^{\mu\nu}}{M^2_V},\ee which leads to the effective
four-fermion Lagrangian \be
\mathcal{L}_{\mathrm{331eff}}=-\fr{1}{2}H^T U^T M^{-2}_V U H.\ee
By isolating in the expression of the currents $(J,J')$ the
relevant electron and quark terms we can identify the coefficients
$C_{1u}$ and $C_{1d}$ in (\ref{leef}). This yields \bea
C_{1u,d}&=&-8\fr{\rho}{c^2_\theta}\left\{\left(c^2_{\theta'}+s^2_{\theta'}\fr{M^2_Z}
{M^2_X}\right)a_e v_{u,d}\right.\crn &&\left.
+\left[\left(1-s^2_{\theta'}(4c^2_W-1)\right)\fr{M^2_Z}{M^2_{Z'}}
+s^2_{\theta'}(4c^2_W-1)\fr{M^2_Z} {M^2_X}\right]a'_e
v'_{u,d}\right.\crn
&&\left.+\left[s^2_{\theta'}\sqrt{4c^2_W-1}\fr{M^2_Z}{M^2_X}
+s_\va\fr{M^2_Z}{M^2_{Z'}}-s^2_{\theta'}
\sqrt{4c^2_W-1}-s_\va\right]\right.\crn && \times\left. (a_e
v'_{u,d}+a'_e v_{u,d})\right\}.\label{c1ud}\eea The SM expressions
for $C_{1u,d}$ can be obtained from (\ref{c1ud}) in the limit
$\om\rightarrow \infty$. Keeping leading order by $\Delta \rho,\
s_\theta^2,\ s_\va,\ M_Z^2/M_{Z'}^2$, and $M_Z^2/M_{X}^2$, the
deviation becomes \bea \de Q_{\mathrm{W}} &=&
16\left\{\fr{1}{16}\left[\left(1+4\fr{s^4_W}{c_{2W}}\right)Z-N\right](\Delta
\rho+s^2_\theta)\right.\crn && -\left.
\fr{1}{16}\left[\left(1-4s^2_W\right)Z-N\right]s^2_{\theta'}
\right. \crn &&\left.+\left[\left(2Z+N\right)a'_e
v'_u+\left(2N+Z\right)a'_e v'_d
\right]\fr{M^2_Z}{M^2_{Z'}}\right.\crn
&&\left.-\left[\left(2Z+N\right)\left(a_e v'_u+ a'_e
v_u\right)+\left(2N+Z\right)\left(a_e v'_d+a'_e
v_d\right)\right]\right.\crn&& \left.
\times\left(s_\va+s^2_{\theta'}\sqrt{4c^2_W-1}\right)
\right\}.\label{detaq}\eea At this first order, only the $Z'$
exchange term, the $W_0-Y_0$, $Z_0-Z'_0-X_0$ mixing terms via
$\theta$, $\va$ and $\theta'$, and the shift $\Delta \rho$ give
contribution to $\de Q_{\mathrm{W}}$. Note the minus sign on the
fourth line in the right-handed side of Eq. (\ref{detaq}) which is
{\it opposite} to that in Ref~\cite{alta}. We  bring attention of
the reader to the fact that, the mixing angles $\xi$
in~\cite{alta} and $\theta$ in~\cite{lalu} are also {\it
opposite}.

With the help of expansions \be s^2_{\theta'}\simeq 4
s^2_\theta/c^2_W,\hs s_\va \simeq
-\fr{7\sqrt{3-4s^2_W}}{2c^2_W}s^2_\theta-\fr{1}{\sqrt{3-4s^2_W}}
\fr{M^2_Z}{M^2_{Z'}},\ee the deviation is  rewritten in the form:
\bea \de Q_{\mathrm{W}} &=&
\left[\left(1+4\fr{s^4_W}{c_{2W}}\right)Z-N\right]\Delta \rho \crn
&& +\left\{\left(3+4t^2_W\right)N+\left(12t^2_W-7+\fr{4}{1-t^4_W}
\right)Z\right.\crn && -8\sqrt{(1+t^2_W)(3-t^2_W)}\left[(2Z+N)(a_e
v'_u+ a'_e v_u)\right.\label{fn}\\ && \left. \left. +(2N+Z)(a_e
v'_d+a'_e v_d)\right]\right\}s^2_\theta
+16\left[\left(2Z+N\right)\right.\crn & & \times \left(a'_e
v'_u+\fr{a_e v'_u+ a'_e v_u}{\sqrt{3-4s^2_W}}\right)
+\left. (2N+Z)\left(a'_e v'_d+\fr{a_e v'_d+a'_e
v_d}{\sqrt{3-4s^2_W}}\right) \right]\fr{M^2_Z}{M^2_{Z'}}.\nn
 \eea

Considering an isotope of the Cesium atom with $Z=55,\ N=78$ and
using $s^2_W=0.2312$, the first term containing the deviation in
the $\rho$ parameter is equal to $-1.1254\ \Delta \rho$ which is
very small  due to $|{\Delta \rho}| < 0.011$ \cite{pdg} (see
also~\cite{il}) and therefore can be ignored \cite{lalu}. In the
case where the first quark generation ($u,d$)  is in triplet, we
have \be \de Q^{\mathrm{tri}}_{\mathrm{W}}=99.1056\ s^2_\theta
+35.5937\ \fr{M^2_Z}{M^2_{Z'}}.\ee Using the experimental value
$\de Q_{\mathrm{W}} = 0.45(48)$ which is $1.1\ \sigma$ away from
the SM predictions \cite{die} and $M_Z=91.1876$ GeV, the mass of
$Z'$ boson and the mixing angles $\theta$, $\va$ are bounded by
\be M_{Z'}
> 564\ \mathrm{GeV},\hs |s_\theta| < 0.097,\hs -0.061<s_\va < 0.\ee
If the first quark generation ($u,d$)  is in antitriplet, the
deviation of the weak charge is given by \be \de
Q^{\mathrm{anti}}_{\mathrm{W}}=498.1060\ s^2_\theta +35.5937\
\fr{M^2_Z}{M^2_{Z'}}. \ee We get the following bounds\be M_{Z'} >
564\ \mathrm{GeV},\hs |s_\theta| < 0.043, \hs -0.018<s_\va < 0.\ee
The evaluation in Ref.\cite{haihiggs} has shown that $s^2_\theta
\ll M^2_Z/M^2_{Z'}$, thus the lower bound on the $Z'$ mass is more
stringent. Indeed, such a lower bound of about 600 GeV has been
obtained from a direct search at the Tevatron \cite{abe}. This
limit on the $Z'$ mass is in agreement with many other bounds on
those in the $\chi$, $\psi$, $\eta$, and $\beta$ models as well as
in the left-right models \cite{pdg}.

Taking into account of the experimental central value $\de
Q_{\mathrm{W}} = 0.45$, we get a typical value on mass for $Z'$
and the $Z-Z'$ mixing angle: \be M_{Z'}\simeq810\ \mathrm{GeV},\hs
s_\va \simeq -0.0088.\ee These values on the mass and the mixing
angle are quite different from those given in Ref.\cite{ponc1}.

\section{ Conclusions}
\label{concl} In this paper we have considered the deviation $\de
Q_W$ of the weak charge $Q_W$ from its SM value in framework of
the recently proposed economical 3-3-1 model. In the considered
model, the deviation $\de Q_W$ gets new contribution from the
mixing among the charged gauge bosons $W$ and $Y$ as well as the
mixing among the real part of the non-Hermitian neutral bilepton
$X$ with the $Z$ and the $Z'$.

Implication to the Cesium atom gives constraints on mass of the
extra neutral gauge boson $Z'$ and on the mixing angles $W-Y$ and
$Z-Z'$. The lower mass limit for $Z'$ is 564 GeV which coincide
with the recent direct search constraints at CDF \cite{abe}, D0
\cite{abazov} and in agreement with various bounds given in
\cite{pdg}. The upper mixing angle limits for $|\theta|$ and
$|\va|$ depend on the quark content. The limits are smaller if the
first quark generation is in antitriplet. Under demanding
explanation on the uncharacteristical heavy of top quark, such
limits are preferred which are quite closer to those given in
\cite{angles}.

If the $W-Y$ mixing is small, and the central value of the
deviation $\de Q_{\mathrm{W}}$ to be $0.45$, the $Z'$ mass and the
$Z-Z'$ mixing angle have the typical values $810$ GeV and
$-0.0088$, respectively. This mass value for the $Z'$ boson is of
relevance for the Tevatron, even before LEP turns on.

{\it Acknowledgments}: This work was supported in part by National
Council for Natural Sciences of Vietnam.

\end{document}